\documentclass[aps,twocolumn,prl,floatfix,showpacs,lengthcheck,citeautoscript,superscriptaddress,nobibnotes,longbibliography]{revtex4-2}
\usepackage{amsmath}
\usepackage{amssymb}
\usepackage{graphicx}
\usepackage{bm}
\usepackage{color,soul}
\usepackage{braket}
\usepackage[dvipsnames,svgnames,table]{xcolor}
\usepackage{times}
\usepackage{MnSymbol}

\usepackage{bbold}
\usepackage{babel}
\usepackage{hyperref}
\usepackage{orcidlink}
\usepackage{lipsum}

\definecolor{Nathanpink}{rgb}{0.94,0.317,0.9607}

\definecolor{titles}{rgb}{0.,0.24,0.51}

\hypersetup{
pdfstartview={FitH},
colorlinks=true,    
linkcolor=NavyBlue, 
citecolor=NavyBlue,   
filecolor=NavyBlue, 
urlcolor=NavyBlue   
}

\begin{document}
\title{Quantized Chern-Simons Axion Coupling in Anomalous Floquet Systems}
\author{Lucila {Peralta Gavensky}\,\orcidlink{0000-0002-5598-7303}}
\email{lucila.peralta.gavensky@ulb.be}
\affiliation{Center for Nonlinear Phenomena and Complex Systems, Universit\'e Libre de Bruxelles, CP 231, Campus Plaine, B-1050 Brussels, Belgium}
\affiliation{International Solvay Institutes, 1050 Brussels, Belgium}
\author{Nathan Goldman\,\orcidlink{0000-0002-0757-7289}}
\affiliation{Center for Nonlinear Phenomena and Complex Systems, Universit\'e Libre de Bruxelles, CP 231, Campus Plaine, B-1050 Brussels, Belgium}
\affiliation{International Solvay Institutes, 1050 Brussels, Belgium}
\affiliation{Laboratoire Kastler Brossel, Coll\`ege de France, CNRS, ENS-Universit\'e PSL, Sorbonne Universit\'e, 11 Place Marcelin Berthelot, 75005 Paris, France}
\author{Gonzalo {Usaj}\,\orcidlink{0000-0002-3044-5778}}
\affiliation{Centro At\'omico Bariloche and Instituto Balseiro, Comisi\'on Nacional de Energ\'ia
Atomica (CNEA)- Universidad Nacional de Cuyo (UNCUYO), 8400 Bariloche, Argentina.}
\affiliation{Instituto de Nanociencia y Nanotecnolog\'ia (INN-Bariloche), Consejo Nacional de Investigaciones Cient\'ificas y Tecnicas (CONICET), Argentina.}
\begin{abstract}
Quantized bulk response functions are hallmark signatures of topological phases, but their manifestation in periodically driven (Floquet) systems is not yet fully established. Here, we show that two-dimensional anomalous Floquet systems exhibit a quantized bulk response encoded in a Chern-Simons axion (CSA) coupling angle $\theta_{\textrm{CS}}^{F} \in 2\pi\mathbb{Z}$, reflecting a topological magnetoelectric effect analogous to that in three-dimensional insulators. The periodic drive introduces an emergent ``photon" dimension, allowing the system to be viewed as a three-dimensional Sambe lattice. Within this framework, cross-correlated responses---namely, photon-space polarization and magnetization density---emerge as physical signatures of the CSA coupling. The CSA angle, constructed from the non-Abelian Berry connection of Floquet states, admits a natural interpretation in terms of the geometry of hybrid Wannier states. These results provide a unified framework linking Floquet band topology to quantized bulk observables.
\end{abstract}
\date{June 25, 2025}
\maketitle
\textbf{\textit{Introduction.---}} Establishing a connection between the topological properties of periodically driven Floquet systems and bulk observables has long posed a significant challenge. One of the key difficulties stems from the fact that the classification of Floquet topological phases builds on the structure of the time-evolution operator itself~\cite{Rudner2013,Nathan2015,Roy2017}, rather than on properties of the system's quantum states. Consequently, a clear geometric interpretation of Floquet topology, rooted in the solutions of the time-dependent Schr\"odinger equation, has not been fully established.

A key conceptual advance was made in Ref.~\cite{Nakagawa2020}, which demonstrated that the topological character of periodically driven systems is encoded in the bulk Floquet modes---specifically, in the nontrivial connectivity of their hybrid Wannier centers defined over time and momentum space. While this framework brought the geometric structure of Floquet eigenstates to the forefront, it did not establish a direct link between the associated topological invariants and experimentally accessible bulk response functions. Interestingly, such a connection has been theoretically established in the case of two-dimensional (2D) driven systems:~the orbital magnetization density of anomalous Floquet systems is quantized according to the topological winding number $N_3[R]$ of Refs.~\cite{Rudner2013,Nathan2015,Roy2017}, or equivalently, to the number of anomalous edge channels connecting different quasienergy zones~\cite{Nathan2017,Glorioso2021,PeraltaGavensky2024}. This result, first derived in the context of Anderson-localized Floquet systems~\cite{Nathan2017}, was recently proven to hold in more generic settings, including clean, translationally invariant lattices~\cite{PeraltaGavensky2024}. 
   \begin{figure}[t]
        \centering
        \includegraphics[width=0.9\columnwidth]{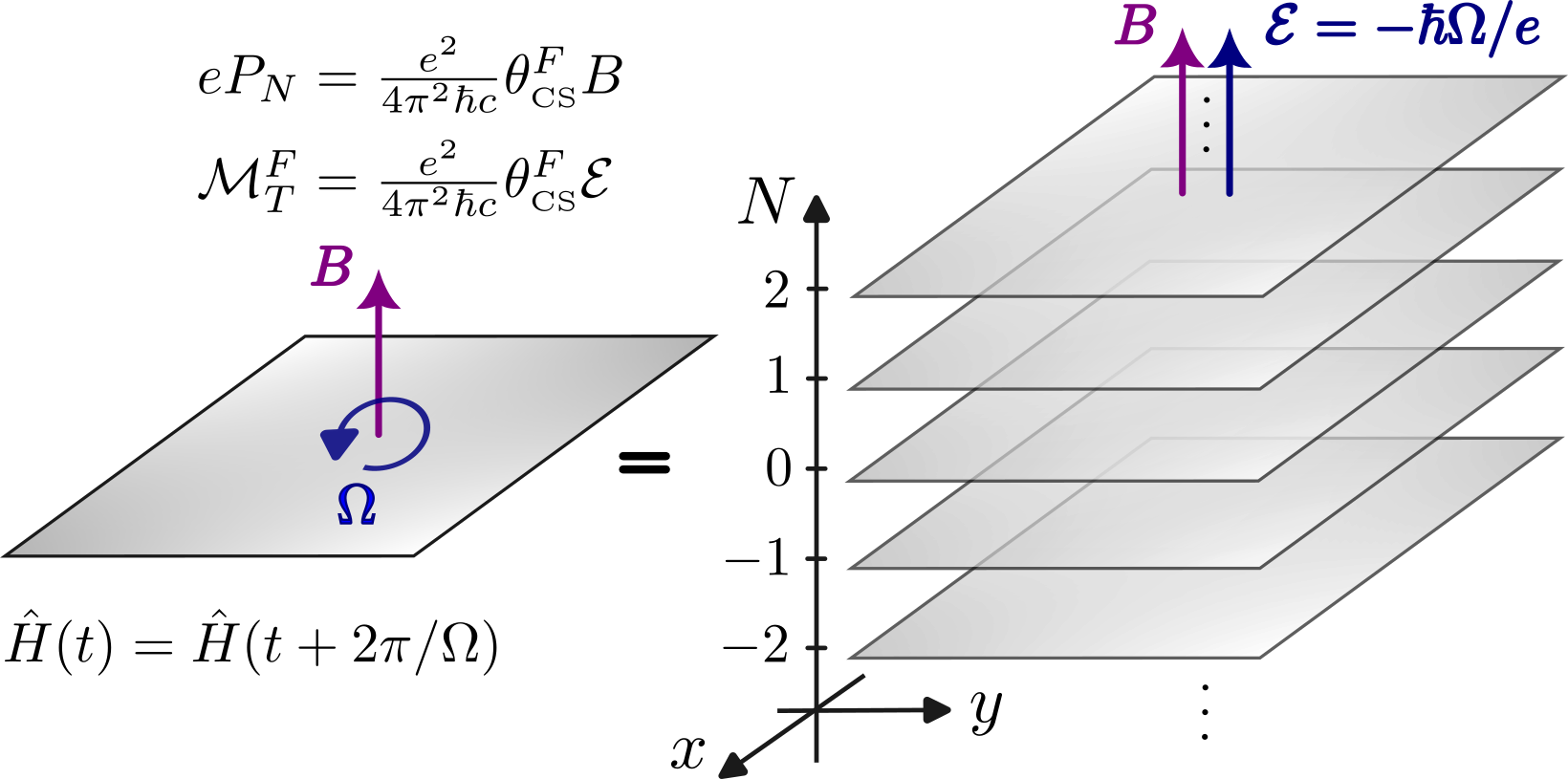}
        \caption{A 2D electronic system driven at frequency $\Omega$, governed by a time-dependent Hamiltonian $\hat{H}(t)$, can be mapped onto a time-independent, 3D Sambe lattice described by the infinite-dimensional Floquet Hamiltonian $\hat{\bm{H}}^F$. In this extended framework, the periodic drive manifests through both the coupling between layers and a static effective electric field $\mathcal{E} = -\hbar\Omega/e$, directed along the emergent photon-number axis. An out-of-plane magnetic field $\bm{B}$ maps to a magnetic field of the same strength along this axis.}
        \label{fig_scheme}
    \end{figure} 

Within a Sambe-space description~\cite{Shirley1965,Sambe1973}, it
 is natural to reinterpret a $D$-dimensional Floquet system as a fictitious $(D+1)$-static system~\cite{GomezLeon2013,Baum2018,Oka2019,Nakagawa2020}, where the extra dimension is provided by a ``photon" degree of freedom. This observation suggests that topological invariants in 2D Floquet systems may correspond to quantized responses, analogous to those found in 3D topological insulators. For instance, it is known that static 3D lattice systems can exhibit a magnetoelectric effect~\cite{Qi2008,Essin2009,Malashevich2010,Shiozaki2013,Sekine2021}, captured by an (emergent) axion electrodynamic Lagrangian $\mathcal{L}\!=\!\theta (e^2/2 \pi h c) \boldsymbol{E} \cdot \boldsymbol{B}$ \cite{Wilczek1987}. However, such a connection between Floquet topology and higher-dimensional topological responses has remained elusive.

 In this work, we demonstrate that 2D anomalous Floquet systems exhibit a quantized Chern-Simons axion (CSA) coupling angle, $\theta_{\textrm{
        CS}}^{F}\!\in\!2\pi \mathbb{Z}$, revealing the fundamental origin of their quantized orbital magnetization in terms of a topological 
magnetoelectric effect. Specifically, we demonstrate that the net number of chiral anomalous edge channels of a 2D Floquet system, i.e.~the Floquet winding number $N_3[R]$,  satisfies the relation
    \begin{equation}
        N_3[R] = \frac{\theta_{\textrm{CS}}^{F}}{2\pi}\,,
        \label{N3=theta}
        \end{equation}
where the CSA coupling angle $\theta_{\textrm{CS}}^{F}$ is constructed from the non-Abelian Berry connection of the Floquet states defined over time and quasimomentum space. This result elucidates the fundamental connection between the anomalous boundary modes of driven 2D systems and a quantized magnetopolarizability along an emergent photon dimension in the 3D Sambe lattice~\cite{Shirley1965,Sambe1973}; see Fig.~\ref{fig_scheme}.  The resulting $\theta_{\textrm{CS}}^F$ term can be physically understood as arising from the interplay between an effective electric field along the photon-axis (intrinsic to Floquet systems) and an applied magnetic-field perturbation $B$ perpendicular to the plane. The CSA coupling manifests through non-trivial cross-correlated responses of the Floquet modes: a photon-domain polarization $P_N$ and the aforementioned orbital magnetization density $\mathcal{M}_T^{F}$. We discuss the implications of these effects and provide a way of expressing the axion coupling $\theta_{\textrm{CS}}^F$ in terms of the geometric properties of the hybrid Wannier states introduced in Ref.~\cite{Nakagawa2020}, closely paralleling known results for 3D static insulators~\cite{Taherinejad2015,Olsen2017}.

\textbf{\textit{From Hilbert space to Sambe space.---}} The evolution operator of a periodically driven system with period $T=2\pi/\Omega$ can be generally expressed as~\cite{Rahav2003,Goldman2014,Eckardt2015,Bukov2015}
\begin{equation}
    \hat{U}(t,t')=\hat{R}(t)e^{-i\hat{H}_{\textrm{eff}}(t-t')/\hbar}\hat{R}^{\dagger}(t')\,,
    \label{UF}
\end{equation}
where $\hat{R}(t)$ is the unitary, $T$-periodic micromotion operator, and $\hat{H}_{\textrm{eff}}$ is the effective (time-independent) Hamiltonian. The eigenvectors of $\hat{H}_{\textrm{eff}}$, denoted as $|u_{a}^{\textrm{eff}}\rangle$, have corresponding eigenvalues $\{\varepsilon_a\}$ spanning the quasienergy spectrum of the Floquet system. As a consequence of Eq.~\eqref{UF}, the solutions of the time-dependent Schr\"odinger equation can always be expressed as $|\psi_a(t)\rangle = e^{-i\varepsilon_a t/\hbar}|u_a(t)\rangle $, where 
\begin{equation}
    |u_a(t)\rangle = \hat{R}(t)|u_{a}^{\textrm{eff}}\rangle\,,
    \label{FM}
\end{equation}
are the $T$-periodic Floquet modes. Importantly, the quasienergies are defined modulo the driving frequency $\hbar\Omega$, leading to a gauge redundancy: the same physical state can be described by
\begin{equation}
    \varepsilon_{as}=\varepsilon_a + s\hbar\Omega,\qquad |u_{as}(t)\rangle = e^{is\Omega t}|u_{a}(t)\rangle\,,\qquad s\in \mathbb{Z}\,.
\end{equation}
Throughout this work, we will take the index $a$ to label solutions within a fixed branch of the quasienergy spectrum, given by the so-called natural Floquet zone (NFZ)~\cite{Nathan2015,PeraltaGavensky2024}. Each of the shifted modes obeys the Floquet eigenvalue equation 
\begin{equation}
    [\hat{H}(t)-i\hbar\partial_t]|u_{as}(t)\rangle = \varepsilon_{as}|u_{as}(t)\rangle\,.
    \label{HF_t}
\end{equation}
To make this structure explicit, we introduce the extended Sambe space $\mathcal{S}=\mathcal{H}\otimes \mathcal{T}$, where $\mathcal{H}$ is the physical Hilbert space and $\mathcal{T}$ is the space of square-integrable $T$-periodic functions~\cite{Shirley1965,Sambe1973}. In this basis, Eq.~\eqref{HF_t} becomes a time-independent eigenvalue problem
\begin{equation}
    \hat{\bm{H}}^{F} |u_{as}\rangle\rangle = [\hat{\bm{H}}-\hbar\Omega \hat{\bm{N}}]|u_{as}\rangle\rangle = \varepsilon_{as}|u_{as}\rangle\rangle\,,
    \label{HF_S}
\end{equation}
where $\hat{\bm{H}}^{}_{nm}=\frac{1}{T}\int_{0}^{T}dt e^{i(n-m)\Omega t}\hat{H}(t)$  encodes the Fourier representation of the original Hamiltonian and  $\hat{\bm{N}}_{nm}=n\delta_{nm}$ is the ``photon-number" operator. The $n$-th block element of the Sambe vector $|u_{as}\rangle\rangle$ is given by
\begin{equation}
    |u_{as}^{(n)}\rangle \equiv \frac{1}{T}\int_{0}^{T}dt e^{in\Omega t}|u_{as}(t)\rangle=|u_a^{(n+s)}\rangle\,,
    \label{HWF}
\end{equation}
where we used the simplified subscript $a 0\!\equiv\!a$ in the last equality.

Equation~\eqref{HF_S} captures the physics of the 3D Sambe lattice, where the original 2D system is extended by a synthetic third dimension corresponding to the photon-number axis, featuring unit lattice spacing (see Fig.~\ref{fig_scheme}). This emergent dimension hosts a uniform effective electric field, $\mathcal{E}\!=\!-\hbar\Omega/e$, that breaks lattice translational symmetry and localizes the Floquet states along the photon-number direction~\cite{Rudner2013,GomezLeon2013,Oka2019}. This field also renders the spectrum of the Floquet Hamiltonian $\hat{\bm{H}}^{F}$ unbounded, reflecting the physics of the Wannier-Stark problem.

\textbf{\textit{Hybrid Wannier representation.---}} When the system exhibits discrete translational symmetry in the two-dimensional plane, the quantum number $a=(\alpha,\bm{k})$, where $\bm{k}=(k_x,k_y)$ is the crystal quasimomentum and $\alpha$ labels the Floquet band.  In this case, Eq.~\eqref{HWF} naturally acquires the interpretation of a hybrid Wannier representation of the Floquet modes $|u_{\alpha s\bm{k}}(t)\rangle$ along the photon-number axis~\cite{Nakagawa2020}. The associated Wannier charge centers (WCCs) define smooth sheets over quasimomentum space and are given by
\begin{equation}
    \overline{N}_{\alpha s}(\bm{k}) = \sum_n n \langle u_{\alpha s \bm{k}}^{(n)}|u_{\alpha s \bm{k}}^{(n)}\rangle = \langle\langle u_{\alpha s\bm{k}}|\hat{\bm{N}}|u_{\alpha s\bm{k}}\rangle\rangle\,.
    \label{wcc}
\end{equation}
Note that the WCC in the $s$-photon cell is displaced relative to the $s=0$ sector---referred to as the \textit{home cell}--- according to $\overline{N}_{\alpha s}(\bm{k}) = \overline{N}_{\alpha}(\bm{k}) - s$, manifesting the inherent gauge freedom in the definition of the Floquet modes. This geometric information is also directly en\-co\-ded in the Floquet eigenenergies, which can be expressed, following Eq.~\eqref{HF_S}, as
\begin{equation}
    \varepsilon_{\alpha s \bm{k}} = \overline{E}_{\alpha\bm{k}}-\hbar\Omega\overline{N}_{\alpha s}(\bm{k})\,,
\end{equation}
with the $s$-independent mean energies being
\begin{equation}
    \overline{E}_{\alpha\bm{k}}\!=\!\langle\langle u_{\alpha s\bm{k}}|\hat{\bm{H}}|u_{\alpha s\bm{k}}\rangle\rangle\!=\!\frac{1}{T}\int_{0}^{T}\!\!\!dt\langle u_{\alpha\bm{k}}(t)|\hat{H}(t)|u_{\alpha\bm{k}}(t)\rangle\,.
\end{equation}
Finally, we note that the WCCs admit an alternative representation in terms of the non-adiabatic Aharonov–Anandan phase accumulated over a driving cycle, obtained by integrating the time-domain Berry connection \cite{Mondragon2019,Nakagawa2020,PeraltaGavensky2024}
\begin{equation}
     \overline{N}_{\alpha s}(\bm{k}) = \frac{1}{2\pi}\int_0^{T} dt\langle u_{\alpha s \bm{k}}(t)|i\partial_t u_{\alpha s\bm{k}}(t)\rangle\,.
     \label{WCC_AA}
\end{equation}

\textbf{\textit{Photon-domain magnetopolarizability.---}}
In previous studies~\cite{Rudner2013,Roy2017}, the number of anomalous edge channels has been related to a higher-order  winding number of the micromotion operator $\hat{R}_{\bm{k}}(t)$ defined as 
\begin{eqnarray}
\notag
    N_3[R] &=& \frac{\epsilon^{zjl}}{8\pi^2}\!\!\int_{0}^{T}\!\!\!\!dt\!\!\int_{\mathrm{BZ}}\!\!d^2k\, \mathrm{tr}\left[\hat{R}^{\dagger}_{\bm{k}}(t)\frac{\partial \hat{R}_{\bm{k}}(t)}{\partial t}\right.\\
    \label{N3}
    &&\left.\hat{R}^{\dagger}_{\bm{k}}(t)\frac{\partial \hat{R}_{\bm{k}}(t)}{\partial k_l}\hat{R}^{\dagger}_{\bm{k}}(t)\frac{\partial \hat{R}_{\bm{k}}(t)}{\partial k_j}\right]\,,
\end{eqnarray}
where the trace runs over all internal degrees of freedom within a unit cell. In Ref.~\cite{PeraltaGavensky2024}, it was shown that this to\-po\-lo\-gi\-cal invariant can be related to the response of a lower-dimensional winding number via
\begin{equation}
   N_3[R] = \frac{\Phi_0}{A_s} \frac{\partial N_1[R]}{\partial B}\,,
   \label{dN1_N3}
\end{equation}
where $A_s$ is the area of the 2D system and $\Phi_0 = hc/e$ is the flux quantum. The quantity $N_1[R]$ denotes the first-order winding number of the micromotion operator, given by
\begin{equation}
  \label{N1}
    N_1[R]=-\frac{i}{2\pi}\int_{0}^{T}dt \mathrm{Tr}[\hat{R}^{\dagger}(t)\partial_t \hat{R}(t)] \in \mathbb{Z}\,,
\end{equation}
where $\mathrm{Tr}[...]$ runs over spatial and internal degrees of freedom. This invariant defines the photon-domain po\-la\-ri\-za\-tion of the Floquet modes within the NFZ, 
\begin{eqnarray}
\notag
    P_N \equiv\frac{N_1[R]}{A_s}= -\frac{1}{A_s}\sum_{a}\langle\langle u_{a}|\hat{\bm{N}}|u_a\rangle\rangle = -\frac{1}{A_s e}\frac{\partial \mathrm{Tr}[\hat{H}_{\textrm{eff}}]}{\partial \mathcal{E}}\,,\\
    \label{PN}
\end{eqnarray}
where, in the last equality, we used the relation $\langle\langle u_{a}|\hat{\bm{N}}|u_{a}\rangle\rangle = -\partial \varepsilon_{a}/\partial \hbar\Omega $, obtained directly via the Hellmann-Feynman theorem in Sambe space. Importantly, the derivative with respect to the effective electric field is evaluated \textit{in the presence} of the field (i.e. at finite $\Omega$). We also note that the polarization is defined modulo an integer number. Indeed, a gauge transformation in the mi\-cro\-mo\-tion operator, $\hat{R}(t)\rightarrow e^{is\Omega t}\hat{R}(t)$, produces a po\-la\-ri\-za\-tion shift $P_{N}\rightarrow P_N + s$.
In crystalline systems, Eq.~\eqref{PN} can be recast in terms of the hybrid WCCs as
\begin{equation}
    P_N = -\sum_{\alpha}\int_{\textrm{BZ}}\frac{d^2k}{(2\pi)^2}\overline{N}_{\alpha}(\bm{k})\,,
\end{equation}
which underscores the geometric nature of $P_N$ and its direct analogy to the polarization along a specific spatial direction in a 3D lattice~\cite{KingSmith1993,Vanderbilt2018}. This identification motivates the interpretation of  $N_3[R]$ in Eq.~\eqref{dN1_N3} as a quantized magnetopolarizability in the synthetic Sambe lattice. Indeed, Eq.~\eqref{dN1_N3} can be recast in a physically transparent form
\begin{equation}
    e \frac{\partial P_N}{\partial B} = \frac{\partial \mathcal{M}_T^{F}}{\partial \mathcal{E}} = \frac{e^2}{2\pi \hbar c}N_{3}[R]\,,
    \label{dPN_dB}
\end{equation}
 which mirrors the quantized magnetoelectric response of 3D topological insulators~\cite{Qi2008,Essin2009,Malashevich2010,Shiozaki2013,Sekine2021}; see also Fig.~\ref{fig_scheme}. Here we used that the total orbital magnetization density is~\cite{PeraltaGavensky2024} 
\begin{equation}
    \mathcal{M}_T^{F}= -\frac{1}{A_s}\frac{\partial \mathrm{Tr}[\hat{H}_{\textrm{eff}}]}{\partial B}\,.
\end{equation}
In light of Eq.~\eqref{dPN_dB}, the identification of the winding number with a CSA coupling emerges as a particularly natural and suggestive ansatz. Following Refs.~\cite{Qi2008,Essin2009}, the Floquet  $\theta_{\textrm{CS}}^{F}$ term can be constructed from the integral of a Chern-Simons 3-form written in terms of the time-periodic Floquet-Bloch modes within the NFZ
    \begin{equation}
    \theta_{\textrm{CS}}^{F} = -\frac{1}{4\pi}\int d^3k \epsilon^{ijl}\mathrm{tr}\left[\mathcal{A}_{i}\partial_j \mathcal{A}_l -i\frac{2}{3}\mathcal{A}_i \mathcal{A}_j \mathcal{A}_l\right]\,,
    \label{theta_CS}
\end{equation}
where $k=(t,\bm{k})$ and
\begin{equation}
    \mathcal{A}_j^{\alpha\beta} =  i \langle u_{\alpha\bm{k}}(t)|\partial_{k_j} u_{\beta\bm{k}}(t)\rangle\,,
\end{equation}
defines the non-Abelian Berry connection in time and quasimomentum space. We note that Eq.~\eqref{theta_CS} can be simplified to 
\begin{eqnarray}
\notag
    \theta_{\textrm{CS}}^{F} &=& -\frac{1}{2\pi}\int_{\textrm{BZ}}d^2 k\int_0^{T}  dt\,\mathrm{tr}\left(\mathcal{A}_y\partial_t \mathcal{A}_x +\mathcal{A}_{0}F_{xy}\right)\,,\\
    &=& -\frac{1}{2\pi}\int_{\textrm{BZ}}d^2 k\int_0^{T}  dt\,\mathrm{tr}\left(\mathcal{A}_y\partial_t \mathcal{A}_x\right)\,,
    \label{theta_CS2}   
\end{eqnarray}
where $F_{xy} =  \mathcal{F}_{xy}-i[\mathcal{A}_x,\mathcal{A}_y]$ is the non-Abelian Berry curvature and $\mathcal{F}_{xy}^{}=\partial_x \mathcal{A}_y - \partial_y\mathcal{A}_x$. In Eq.~\eqref{theta_CS2}, we have used that $F_{xy}$ vanishes identically, as the connections are defined within the full manifold of Floquet states within the NFZ. With these expressions at hand, and upon substituting Eq.~\eqref{FM} into Eq.~\eqref{N3}, one finally obtains the key result announced in Eq.~\eqref{N3=theta}, formally establishing the identification of the Floquet winding number with a quantized 3D theta term: $N_3[R] = \theta_{\textrm{CS}}^F/2\pi$. This simple yet striking relation provides the fundamental origin of the quantization of $\mathcal{M}_T^{F}$ in units of $\hbar \Omega/\Phi_0$ in anomalous Floquet phases~\cite{Nathan2017,Glorioso2021,PeraltaGavensky2024}.

Interestingly, the hybrid Wannier representation of Floquet states, allows for yet an alternative expression for the CSA coupling. The Berry connections in the Bloch representation can be decomposed as
\begin{equation}
    \mathcal{A}_{x(y)}^{\alpha\beta}=\sum_l e^{-il\Omega t} \mathcal{A}_{x(y)}^{\alpha0;\beta l}\, ,
\end{equation}
with
\begin{equation}
    \mathcal{A}_{x(y)}^{\alpha0;\beta l} = \langle\langle u_{\alpha\bm{k}}|i\partial_{k_{x(y)}} u_{\beta l\bm{k}}\rangle\rangle\, ,
\end{equation}
being the connections defined on the Floquet Wannier sheets. Performing the time integration in Eq.~\eqref{theta_CS2} explicitly, we obtain
\begin{subequations}
\begin{align}
   \theta^{F}_{\textrm{CS}} =&\,\theta^{}_{N\mathcal{F}}+\theta^{}_{\Delta xy}\,,\\
    \theta^{}_{N\mathcal{F}}=&-\sum_{\alpha}\int_{\textrm{BZ}}d^2k \overline{N}_{\alpha}\mathcal{F}_{xy}^{\alpha 0;\alpha 0}\,,\\
    \label{theta_Dxy}
    \theta^{}_{\Delta xy} =& -i\sum_{\alpha\beta s}\int_{\textrm{BZ}}d^2k(\overline{N}_{\beta s}-\overline{N}_{\alpha})\mathcal{A}_{x}^{\alpha 0;\beta s}\mathcal{A}_y^{\beta s;\alpha 0}\,,
\end{align}
\label{CSA_HWF_global}
\end{subequations}
where $\mathcal{F}_{xy}^{\alpha 0;\alpha 0} = \partial_{k_x}\mathcal{A}_y^{\alpha 0;\alpha 0}-\partial_{k_y}\mathcal{A}_x^{\alpha 0;\alpha 0}$ is the Abelian Berry curvature of the hybrid Floquet-Wannier state labeled by $(\alpha,0)$, and where we have omitted the explicit $\bm{k}$-dependence of the integrands for brevity. Remarkably, the decomposition in Eq.~\eqref{CSA_HWF_global} closely parallels that found for static 3D crystals~\cite{Taherinejad2015,Olsen2017}, with the WCCs along the photon-axis playing the role of WCCs along the $z$-direction.  We also note that the $\theta_{N\mathcal{F}}$ contribution can be regarded as a Berry curvature dipole term along the photon direction.

The overall correspondence between our expressions and the ones of Refs.~\cite{Taherinejad2015,Olsen2017} not only reinforces the analogy between anomalous 2D Floquet systems and static 3D topological insulators, but also provides a purely geometric formulation of the associated Floquet invariant. While the present derivation builds on a Floquet-Bloch representation, we point out that a real-space description~\cite{PeraltaGavensky2024,Olsen2017} would equally apply and lead to similar results in the case of non-translationally-invariant systems.

\textbf{\textit{Illustrative example.---}} To demonstrate the applicability of the above framework, we turn to the Kitagawa-type model~\cite{Kitagawa2010} studied in Ref.~\cite{PeraltaGavensky2024}. The system is described by a tight-binding Hamiltonian on a honeycomb lattice,
\begin{eqnarray}
\label{H_Kitagawa}
    \hat{H}(t) &=& \sum_{\bm{R}\in \mathrm{A}}\sum_{\nu=1}^{3}\left(J_{\nu}(t)\hat{c}^{\dagger}_{\bm{R}}\hat{c}^{}_{\bm{R}+\bm{\delta}_\nu}+ {\rm h.c.}\right)\\
    \notag
    &+& \Delta\left(\sum_{\bm{R}\in\mathrm{A}}\hat{c}^{\dagger}_{\bm{R}}\hat{c}^{}_{\bm{R}} -\sum_{\bm{R}\in\mathrm{B}}\hat{c}^{\dagger}_{\bm{R}}\hat{c}^{}_{\bm{R}}\right)\,,  
\end{eqnarray}
where $\bm{\delta}_1\!=\!(0,a_0)$, $\bm{\delta}_2\!=\!(-\sqrt{3}a_0/2,-a_0/2)$ and $\bm{\delta}_3\!=\!(\sqrt{3}a_0/2,-a_0/2)$, with $a_0$ being the distance between neighboring sites. The hopping elements are modulated in time in a chiral manner according to
\begin{equation}
    J_{\nu}(t) = J\, e^{\lambda \cos(\Omega t + \varphi_{\nu})}\,,
    \label{drv_prot}
\end{equation}
where $\varphi_{\nu}\!=\!-\frac{4\pi}{3\sqrt{3}a_0}\bm{\delta}_\nu\cdot \bm{e_x}$, with $\bm{e_x}\!=\!(1,0)$, and $\lambda$ is a dimensionless driving strength. The total bandwidth of the corresponding time-averaged Bloch-Hamiltonian is given by $W=2\sqrt{\Delta^2 + 9J^2\mathcal{I}^2_0(\lambda)}$, where $\mathcal{I}_0(\lambda)$ is the modified Bessel function of the first kind. In the high-frequency regime, where $\hbar\Omega \gg W$, the dynamics of the driven Floquet system is well-captured by a local effective Hamiltonian $\hat{H}_{\textrm{eff}}$, which takes the form of a Haldane-type model. In contrast, when $\hbar\Omega \simeq W$, a resonance occurs, signaled by a gap closing at the Floquet zone edge. This leads to the emergence of an anomalous Floquet topological phase, where all bulk bands have zero Chern number, yet a single chiral edge mode traverses all quasienergy gaps---see Ref.~\cite{PeraltaGavensky2024} for details.

\begin{figure}[t]
    \centering
    \includegraphics[width=\columnwidth]{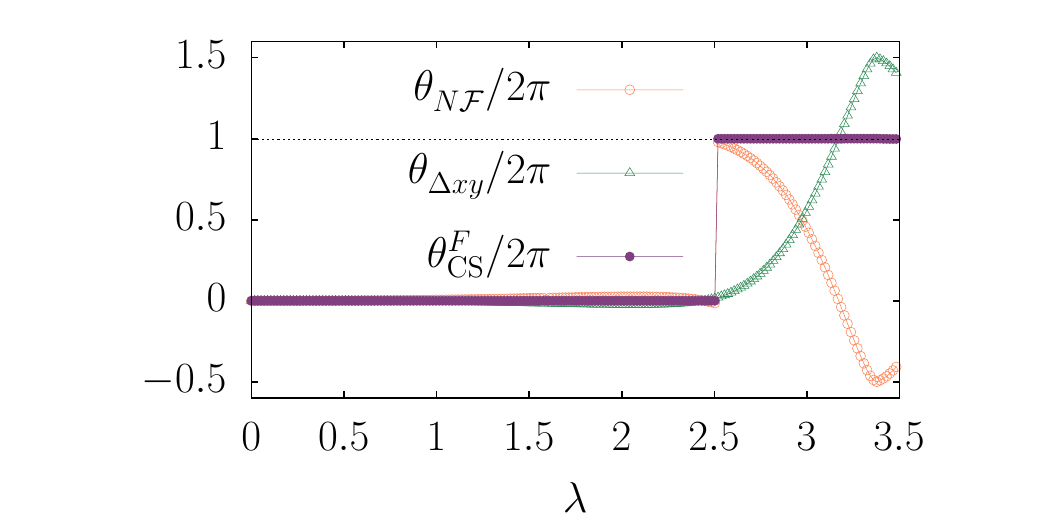}
    \caption{Floquet Chern-Simons axion coupling angle $\theta_{\textrm{CS}}^F$ of the model defined in Eq.~\eqref{H_Kitagawa}, plotted as a function of the dimensionless driving strength $\lambda$ for $\hbar\Omega/J = 20$ and $\Delta/J = 0.5$. The total CSA coupling is shown along with its two separate contributions, $\theta_{N\mathcal{F}}$ and $\theta_{\Delta x y}$, as defined in Eq.~\eqref{CSA_HWF_global}.}
    \label{fig_CSA_wannier}
\end{figure}

\begin{figure}[t]
    \centering
    \includegraphics[width=\columnwidth]{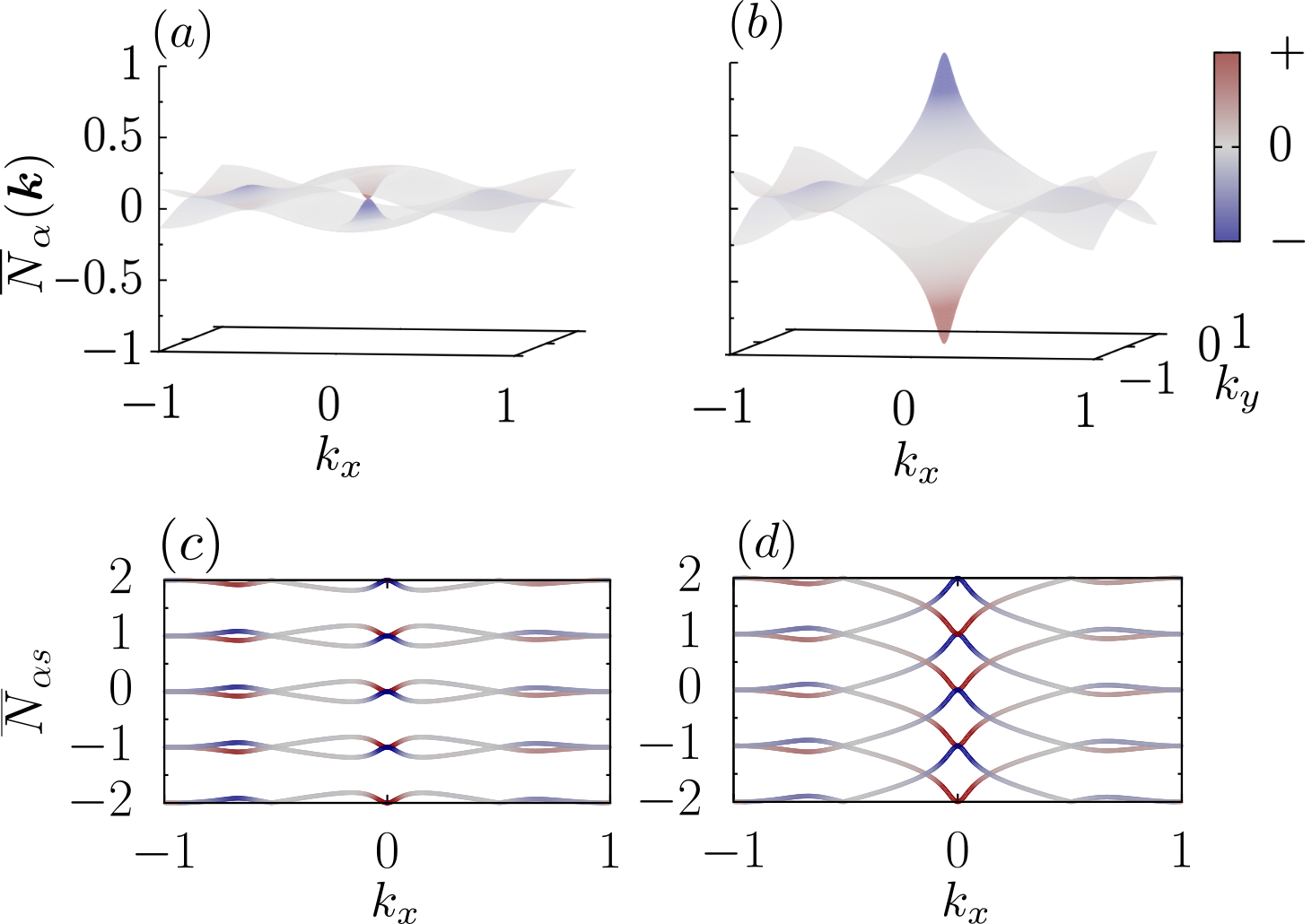}
    \caption{Panels $(a)$ and $(b)$ show the dispersion of the WCC sheets in the home cell, $\overline{N}_{\alpha}(\bm{k})$, for $\hbar\Omega/J=20$ and $\Delta/J=0.5$ in the Kitagawa model defined in Eq.~\eqref{H_Kitagawa}. The quasimomentum components are given in units of $2\pi/\sqrt{3}a_0$ for $k_x$ and $\pi/3 a_0$ for $k_y$. Panel $(a)$ corresponds to $\lambda=2.4$, while panel $(b)$ lies in the anomalous regime with $\lambda=2.6$. Red and blue colors show, respectively, positive and negative values of Berry curvature $\mathcal{F}_{xy}^{\alpha0;\alpha 0}(\bm{k})$ on the sheets. Panels $(c)$ and $(d)$ show  the dispersion of the WCC sheets and their shifted replicas along $k_y=0$.}
    \label{fig_WCC_sheets}
\end{figure}

The Floquet theta term $\theta_{\textrm{CS}}^{F}$ of this tight-binding model is plotted in Fig.~\ref{fig_CSA_wannier} as a function of the parameter $\lambda$, together with its two contributions $\theta_{N\mathcal{F}}$ and $\theta_{\Delta x y}$; see Eq.~\eqref{CSA_HWF_global}. For the chosen parameters, the transition to the anomalous Floquet phase with $N_3[R]=1$ occurs at $\lambda \simeq 2.5$, marked by a quantized jump of $2\pi$ in $\theta_{\textrm{CS}}^{F}$. Interestingly, it is the Berry curvature dipole term, $\theta_{N\mathcal{F}}$, that is responsible for the discontinuous jump at the transition. This is consistent with the quasienergy band inversion occurring at the Floquet zone edge, which in turn triggers an abrupt redistribution of the Berry curvature $\mathcal{F}_{xy}^{\alpha 0;\alpha 0} (\bm{k})$ of the Wannier sheets in the home-cell. This effect is illustrated in Fig.~\ref{fig_WCC_sheets}, where we show the dispersion of the WCC sheets across the Brillouin zone, $\overline{N}_{\alpha}(\bm{k})$, just before [Fig.~\ref{fig_WCC_sheets}$(a)$] and after [Fig.~\ref{fig_WCC_sheets}$(b)$] the transition into the anomalous regime. Red and blue colors indicate regions of positive and negative Berry curvature, respectively. In the anomalous phase, the home-cell WCCs attain their extrema at the $\Gamma$ ($\bm{k}=\bm{0}$) point, accompanied by an abrupt sign reversal of the Berry curvature relative to the non-anomalous regime. This sharp redistribution is directly responsible for the jump in the dipole contribution  $\theta_{N\mathcal{F}}$ term. The corresponding dispersion of the WCC sheets and their periodic images along $k_y=0$ is plotted in Figs.~\ref{fig_WCC_sheets}$(c)$ and $(d)$, revealing how the sheets begin to overlap in the anomalous phase.

\textbf{\textit{Discussion \& Perspectives.---}} In this work, we established a formal connection between the quantized bulk responses of anomalous 2D Floquet phases and the manifestation of a non-trivial 3D magnetoelectric effect along an emergent photon dimension, characterized by a quantized CSA coupling angle $\theta_{\textrm{CS}}^{F}$. By employing a Wannier representation of the Floquet states in the photon domain, we arrived at a purely geometric and eigenstate-based formulation of the Floquet invariant $N_3[R]$, expressed entirely in terms of Berry potentials and Wannier charge centers. We anticipate that this approach can be naturally extended to Floquet systems in other spatial dimensions and symmetry classes, providing a unified framework to understand the bulk topological responses of periodically driven systems.

From an experimental standpoint, it would be highly interesting to probe the cross-correlated photon-space polarization and magnetization densities. A promising platform is provided by artificial Floquet electrical circuits~\cite{Stegmaier2024,Zhang2025}, which not only offer direct access to the frequency-space structure of Floquet eigenstates, but also enable the engineering of boundaries along the emergent synthetic dimension; an alternative approach involves employing memory kernels, as proposed in Ref.~\cite{Baum2018}. This unique capability opens the door to experimental investigations of bulk-surface phenomena within the Sambe lattice framework.

From a theoretical perspective, it would be compelling to investigate whether the predicted bulk responses can be derived from an effective topological field theory formulated in Sambe space~\cite{Glorioso2021}, where the coupling between the synthetic electric field $\mathcal{E}$ and the applied magnetic field $B$ could naturally follow from the underlying equations of motion. We also note that the effective field $\mathcal{E}$ could be made dynamical by introducing a slow time dependence in the driving frequency. Additionally, engineering spatial variations in the driving parameters could serve as an extra tuning knob~\cite{Kishony2025}. Whether a time and position dependent axion angle $\theta_{\textrm{CS}}^{F}(\bm{r},\tau)$---with $\tau$ denoting the slow time-scale---can give rise to non-trivial axion-like electrodynamics remains an intriguing open question.\\

\begin{acknowledgments}
\textbf{\textit{Acknowledgments.---}} We acknowledge David Vanderbilt for pointing out his works on the CSA coupling in 3D topological insulators and Shinsei Ryu for useful discussions. This research was financially supported by the FRS-FNRS (Belgium), the ERC Grant LATIS and the EOS project CHEQS. LPG  acknowledges support
provided by the L’Or\'eal-UNESCO for Women in Science Programme. GU acknowledges financial support from the ANPCyT-FONCyT (Argentina) under grant PICT 2019-0371, SeCyT-UNCuyo grant 06/C053-T1 and the FRS-FNRS for a scientific research stay grant 2024/V 6/5/012.
\end{acknowledgments}
%


\begin{thebibliography}{31}%
\makeatletter
\providecommand \@ifxundefined [1]{%
 \@ifx{#1\undefined}
}%
\providecommand \@ifnum [1]{%
 \ifnum #1\expandafter \@firstoftwo
 \else \expandafter \@secondoftwo
 \fi
}%
\providecommand \@ifx [1]{%
 \ifx #1\expandafter \@firstoftwo
 \else \expandafter \@secondoftwo
 \fi
}%
\providecommand \natexlab [1]{#1}%
\providecommand \enquote  [1]{``#1''}%
\providecommand \bibnamefont  [1]{#1}%
\providecommand \bibfnamefont [1]{#1}%
\providecommand \citenamefont [1]{#1}%
\providecommand \href@noop [0]{\@secondoftwo}%
\providecommand \href [0]{\begingroup \@sanitize@url \@href}%
\providecommand \@href[1]{\@@startlink{#1}\@@href}%
\providecommand \@@href[1]{\endgroup#1\@@endlink}%
\providecommand \@sanitize@url [0]{\catcode `\\12\catcode `\$12\catcode
  `\&12\catcode `\#12\catcode `\^12\catcode `\_12\catcode `\%12\relax}%
\providecommand \@@startlink[1]{}%
\providecommand \@@endlink[0]{}%
\providecommand \url  [0]{\begingroup\@sanitize@url \@url }%
\providecommand \@url [1]{\endgroup\@href {#1}{\urlprefix }}%
\providecommand \urlprefix  [0]{URL }%
\providecommand \Eprint [0]{\href }%
\providecommand \doibase [0]{https://doi.org/}%
\providecommand \selectlanguage [0]{\@gobble}%
\providecommand \bibinfo  [0]{\@secondoftwo}%
\providecommand \bibfield  [0]{\@secondoftwo}%
\providecommand \translation [1]{[#1]}%
\providecommand \BibitemOpen [0]{}%
\providecommand \bibitemStop [0]{}%
\providecommand \bibitemNoStop [0]{.\EOS\space}%
\providecommand \EOS [0]{\spacefactor3000\relax}%
\providecommand \BibitemShut  [1]{\csname bibitem#1\endcsname}%
\let\auto@bib@innerbib\@empty
\bibitem [{\citenamefont {Rudner}\ \emph {et~al.}(2013)\citenamefont {Rudner},
  \citenamefont {Lindner}, \citenamefont {Berg},\ and\ \citenamefont
  {Levin}}]{Rudner2013}%
  \BibitemOpen
  \bibfield  {author} {\bibinfo {author} {\bibfnamefont {M.~S.}\ \bibnamefont
  {Rudner}}, \bibinfo {author} {\bibfnamefont {N.~H.}\ \bibnamefont {Lindner}},
  \bibinfo {author} {\bibfnamefont {E.}~\bibnamefont {Berg}},\ and\ \bibinfo
  {author} {\bibfnamefont {M.}~\bibnamefont {Levin}},\ }\bibfield  {title}
  {\bibinfo {title} {{Anomalous Edge States and the Bulk-Edge Correspondence
  for Periodically Driven Two-Dimensional Systems}},\ }\href
  {https://doi.org/10.1103/PhysRevX.3.031005} {\bibfield  {journal} {\bibinfo
  {journal} {Phys. Rev. X}\ }\textbf {\bibinfo {volume} {3}},\ \bibinfo {pages}
  {031005} (\bibinfo {year} {2013})}\BibitemShut {NoStop}%
\bibitem [{\citenamefont {Nathan}\ and\ \citenamefont
  {Rudner}(2015)}]{Nathan2015}%
  \BibitemOpen
  \bibfield  {author} {\bibinfo {author} {\bibfnamefont {F.}~\bibnamefont
  {Nathan}}\ and\ \bibinfo {author} {\bibfnamefont {M.~S.}\ \bibnamefont
  {Rudner}},\ }\bibfield  {title} {\bibinfo {title} {{Topological singularities
  and the general classification of Floquet–Bloch systems}},\ }\href
  {https://doi.org/10.1088/1367-2630/17/12/125014} {\bibfield  {journal}
  {\bibinfo  {journal} {New Journal of Physics}\ }\textbf {\bibinfo {volume}
  {17}},\ \bibinfo {pages} {125014} (\bibinfo {year} {2015})}\BibitemShut
  {NoStop}%
\bibitem [{\citenamefont {Roy}\ and\ \citenamefont {Harper}(2017)}]{Roy2017}%
  \BibitemOpen
  \bibfield  {author} {\bibinfo {author} {\bibfnamefont {R.}~\bibnamefont
  {Roy}}\ and\ \bibinfo {author} {\bibfnamefont {F.}~\bibnamefont {Harper}},\
  }\bibfield  {title} {\bibinfo {title} {{Periodic table for Floquet
  topological insulators}},\ }\href
  {https://doi.org/10.1103/PhysRevB.96.155118} {\bibfield  {journal} {\bibinfo
  {journal} {Phys. Rev. B}\ }\textbf {\bibinfo {volume} {96}},\ \bibinfo
  {pages} {155118} (\bibinfo {year} {2017})}\BibitemShut {NoStop}%
\bibitem [{\citenamefont {Nakagawa}\ \emph {et~al.}(2020)\citenamefont
  {Nakagawa}, \citenamefont {Slager}, \citenamefont {Higashikawa},\ and\
  \citenamefont {Oka}}]{Nakagawa2020}%
  \BibitemOpen
  \bibfield  {author} {\bibinfo {author} {\bibfnamefont {M.}~\bibnamefont
  {Nakagawa}}, \bibinfo {author} {\bibfnamefont {R.-J.}\ \bibnamefont
  {Slager}}, \bibinfo {author} {\bibfnamefont {S.}~\bibnamefont
  {Higashikawa}},\ and\ \bibinfo {author} {\bibfnamefont {T.}~\bibnamefont
  {Oka}},\ }\bibfield  {title} {\bibinfo {title} {{Wannier representation of
  Floquet topological states}},\ }\href
  {https://doi.org/10.1103/PhysRevB.101.075108} {\bibfield  {journal} {\bibinfo
   {journal} {Phys. Rev. B}\ }\textbf {\bibinfo {volume} {101}},\ \bibinfo
  {pages} {075108} (\bibinfo {year} {2020})}\BibitemShut {NoStop}%
\bibitem [{\citenamefont {Nathan}\ \emph {et~al.}(2017)\citenamefont {Nathan},
  \citenamefont {Rudner}, \citenamefont {Lindner}, \citenamefont {Berg},\ and\
  \citenamefont {Refael}}]{Nathan2017}%
  \BibitemOpen
  \bibfield  {author} {\bibinfo {author} {\bibfnamefont {F.}~\bibnamefont
  {Nathan}}, \bibinfo {author} {\bibfnamefont {M.~S.}\ \bibnamefont {Rudner}},
  \bibinfo {author} {\bibfnamefont {N.~H.}\ \bibnamefont {Lindner}}, \bibinfo
  {author} {\bibfnamefont {E.}~\bibnamefont {Berg}},\ and\ \bibinfo {author}
  {\bibfnamefont {G.}~\bibnamefont {Refael}},\ }\bibfield  {title} {\bibinfo
  {title} {{Quantized Magnetization Density in Periodically Driven Systems}},\
  }\href {https://doi.org/10.1103/PhysRevLett.119.186801} {\bibfield  {journal}
  {\bibinfo  {journal} {Phys. Rev. Lett.}\ }\textbf {\bibinfo {volume} {119}},\
  \bibinfo {pages} {186801} (\bibinfo {year} {2017})}\BibitemShut {NoStop}%
\bibitem [{\citenamefont {Glorioso}\ \emph {et~al.}(2021)\citenamefont
  {Glorioso}, \citenamefont {Gromov},\ and\ \citenamefont
  {Ryu}}]{Glorioso2021}%
  \BibitemOpen
  \bibfield  {author} {\bibinfo {author} {\bibfnamefont {P.}~\bibnamefont
  {Glorioso}}, \bibinfo {author} {\bibfnamefont {A.}~\bibnamefont {Gromov}},\
  and\ \bibinfo {author} {\bibfnamefont {S.}~\bibnamefont {Ryu}},\ }\bibfield
  {title} {\bibinfo {title} {{Effective response theory for Floquet topological
  systems}},\ }\href {https://doi.org/10.1103/PhysRevResearch.3.013117}
  {\bibfield  {journal} {\bibinfo  {journal} {Phys. Rev. Res.}\ }\textbf
  {\bibinfo {volume} {3}},\ \bibinfo {pages} {013117} (\bibinfo {year}
  {2021})}\BibitemShut {NoStop}%
\bibitem [{\citenamefont {Gavensky}\ \emph {et~al.}(2024)\citenamefont
  {Gavensky}, \citenamefont {Usaj},\ and\ \citenamefont
  {Goldman}}]{PeraltaGavensky2024}%
  \BibitemOpen
  \bibfield  {author} {\bibinfo {author} {\bibfnamefont {L.~P.}\ \bibnamefont
  {Gavensky}}, \bibinfo {author} {\bibfnamefont {G.}~\bibnamefont {Usaj}},\
  and\ \bibinfo {author} {\bibfnamefont {N.}~\bibnamefont {Goldman}},\
  }\bibfield  {title} {\bibinfo {title} {{The St\v{r}eda Formula for Floquet
  Systems: Topological Invariants and Quantized Anomalies from Ces\`aro
  Summation}},\ }\Eprint {https://arxiv.org/abs/2408.13576} {arXiv:2408.13576}
  (\bibinfo {year} {2024})\BibitemShut {NoStop}%
\bibitem [{\citenamefont {Shirley}(1965)}]{Shirley1965}%
  \BibitemOpen
  \bibfield  {author} {\bibinfo {author} {\bibfnamefont {J.~H.}\ \bibnamefont
  {Shirley}},\ }\bibfield  {title} {\bibinfo {title} {{Solution of the
  Schr\"odinger Equation with a Hamiltonian Periodic in Time}},\ }\href
  {https://doi.org/10.1103/PhysRev.138.B979} {\bibfield  {journal} {\bibinfo
  {journal} {Phys. Rev.}\ }\textbf {\bibinfo {volume} {138}},\ \bibinfo {pages}
  {B979} (\bibinfo {year} {1965})}\BibitemShut {NoStop}%
\bibitem [{\citenamefont {Sambe}(1973)}]{Sambe1973}%
  \BibitemOpen
  \bibfield  {author} {\bibinfo {author} {\bibfnamefont {H.}~\bibnamefont
  {Sambe}},\ }\bibfield  {title} {\bibinfo {title} {{Steady States and
  Quasienergies of a Quantum-Mechanical System in an Oscillating Field}},\
  }\href {https://doi.org/10.1103/PhysRevA.7.2203} {\bibfield  {journal}
  {\bibinfo  {journal} {Phys. Rev. A}\ }\textbf {\bibinfo {volume} {7}},\
  \bibinfo {pages} {2203} (\bibinfo {year} {1973})}\BibitemShut {NoStop}%
\bibitem [{\citenamefont {{G\'omez-Le\'on, A. and Platero,
  G.}}(2013)}]{GomezLeon2013}%
  \BibitemOpen
  \bibfield  {author} {\bibinfo {author} {\bibnamefont {{G\'omez-Le\'on, A. and
  Platero, G.}}},\ }\bibfield  {title} {\bibinfo {title} {{Floquet-Bloch Theory
  and Topology in Periodically Driven Lattices}},\ }\href
  {https://doi.org/10.1103/PhysRevLett.110.200403} {\bibfield  {journal}
  {\bibinfo  {journal} {Phys. Rev. Lett.}\ }\textbf {\bibinfo {volume} {110}},\
  \bibinfo {pages} {200403} (\bibinfo {year} {2013})}\BibitemShut {NoStop}%
\bibitem [{\citenamefont {Baum}\ and\ \citenamefont {Refael}(2018)}]{Baum2018}%
  \BibitemOpen
  \bibfield  {author} {\bibinfo {author} {\bibfnamefont {Y.}~\bibnamefont
  {Baum}}\ and\ \bibinfo {author} {\bibfnamefont {G.}~\bibnamefont {Refael}},\
  }\bibfield  {title} {\bibinfo {title} {{Setting Boundaries with Memory:
  Generation of Topological Boundary States in Floquet-Induced Synthetic
  Crystals}},\ }\href {https://doi.org/10.1103/PhysRevLett.120.106402}
  {\bibfield  {journal} {\bibinfo  {journal} {Phys. Rev. Lett.}\ }\textbf
  {\bibinfo {volume} {120}},\ \bibinfo {pages} {106402} (\bibinfo {year}
  {2018})}\BibitemShut {NoStop}%
\bibitem [{\citenamefont {Oka}\ and\ \citenamefont {Kitamura}(2019)}]{Oka2019}%
  \BibitemOpen
  \bibfield  {author} {\bibinfo {author} {\bibfnamefont {T.}~\bibnamefont
  {Oka}}\ and\ \bibinfo {author} {\bibfnamefont {S.}~\bibnamefont {Kitamura}},\
  }\bibfield  {title} {\bibinfo {title} {{Floquet Engineering of Quantum
  Materials}},\ }\href
  {https://doi.org/10.1146/annurev-conmatphys-031218-013423} {\bibfield
  {journal} {\bibinfo  {journal} {Annual Review of Condensed Matter Physics}\
  }\textbf {\bibinfo {volume} {10}},\ \bibinfo {pages} {387–408} (\bibinfo
  {year} {2019})}\BibitemShut {NoStop}%
\bibitem [{\citenamefont {Qi}\ \emph {et~al.}(2008)\citenamefont {Qi},
  \citenamefont {Hughes},\ and\ \citenamefont {Zhang}}]{Qi2008}%
  \BibitemOpen
  \bibfield  {author} {\bibinfo {author} {\bibfnamefont {X.-L.}\ \bibnamefont
  {Qi}}, \bibinfo {author} {\bibfnamefont {T.~L.}\ \bibnamefont {Hughes}},\
  and\ \bibinfo {author} {\bibfnamefont {S.-C.}\ \bibnamefont {Zhang}},\
  }\bibfield  {title} {\bibinfo {title} {{Topological field theory of
  time-reversal invariant insulators}},\ }\href
  {https://doi.org/10.1103/PhysRevB.78.195424} {\bibfield  {journal} {\bibinfo
  {journal} {Phys. Rev. B}\ }\textbf {\bibinfo {volume} {78}},\ \bibinfo
  {pages} {195424} (\bibinfo {year} {2008})}\BibitemShut {NoStop}%
\bibitem [{\citenamefont {Essin}\ \emph {et~al.}(2009)\citenamefont {Essin},
  \citenamefont {Moore},\ and\ \citenamefont {Vanderbilt}}]{Essin2009}%
  \BibitemOpen
  \bibfield  {author} {\bibinfo {author} {\bibfnamefont {A.~M.}\ \bibnamefont
  {Essin}}, \bibinfo {author} {\bibfnamefont {J.~E.}\ \bibnamefont {Moore}},\
  and\ \bibinfo {author} {\bibfnamefont {D.}~\bibnamefont {Vanderbilt}},\
  }\bibfield  {title} {\bibinfo {title} {{Magnetoelectric Polarizability and
  Axion Electrodynamics in Crystalline Insulators}},\ }\href
  {https://doi.org/10.1103/PhysRevLett.102.146805} {\bibfield  {journal}
  {\bibinfo  {journal} {Phys. Rev. Lett.}\ }\textbf {\bibinfo {volume} {102}},\
  \bibinfo {pages} {146805} (\bibinfo {year} {2009})}\BibitemShut {NoStop}%
\bibitem [{\citenamefont {Malashevich}\ \emph {et~al.}(2010)\citenamefont
  {Malashevich}, \citenamefont {Souza}, \citenamefont {Coh},\ and\
  \citenamefont {Vanderbilt}}]{Malashevich2010}%
  \BibitemOpen
  \bibfield  {author} {\bibinfo {author} {\bibfnamefont {A.}~\bibnamefont
  {Malashevich}}, \bibinfo {author} {\bibfnamefont {I.}~\bibnamefont {Souza}},
  \bibinfo {author} {\bibfnamefont {S.}~\bibnamefont {Coh}},\ and\ \bibinfo
  {author} {\bibfnamefont {D.}~\bibnamefont {Vanderbilt}},\ }\bibfield  {title}
  {\bibinfo {title} {{Theory of orbital magnetoelectric response}},\ }\href
  {https://doi.org/10.1088/1367-2630/12/5/053032} {\bibfield  {journal}
  {\bibinfo  {journal} {New Journal of Physics}\ }\textbf {\bibinfo {volume}
  {12}},\ \bibinfo {pages} {053032} (\bibinfo {year} {2010})}\BibitemShut
  {NoStop}%
\bibitem [{\citenamefont {Shiozaki}\ and\ \citenamefont
  {Fujimoto}(2013)}]{Shiozaki2013}%
  \BibitemOpen
  \bibfield  {author} {\bibinfo {author} {\bibfnamefont {K.}~\bibnamefont
  {Shiozaki}}\ and\ \bibinfo {author} {\bibfnamefont {S.}~\bibnamefont
  {Fujimoto}},\ }\bibfield  {title} {\bibinfo {title} {{Electromagnetic and
  Thermal Responses of $Z$ Topological Insulators and Superconductors in Odd
  Spatial Dimensions}},\ }\href
  {https://doi.org/10.1103/PhysRevLett.110.076804} {\bibfield  {journal}
  {\bibinfo  {journal} {Phys. Rev. Lett.}\ }\textbf {\bibinfo {volume} {110}},\
  \bibinfo {pages} {076804} (\bibinfo {year} {2013})}\BibitemShut {NoStop}%
\bibitem [{\citenamefont {Sekine}\ and\ \citenamefont
  {Nomura}(2021)}]{Sekine2021}%
  \BibitemOpen
  \bibfield  {author} {\bibinfo {author} {\bibfnamefont {A.}~\bibnamefont
  {Sekine}}\ and\ \bibinfo {author} {\bibfnamefont {K.}~\bibnamefont
  {Nomura}},\ }\bibfield  {title} {\bibinfo {title} {{Axion electrodynamics in
  topological materials}},\ }\bibfield  {journal} {\bibinfo  {journal} {Journal
  of Applied Physics}\ }\textbf {\bibinfo {volume} {129}},\ \href
  {https://doi.org/10.1063/5.0038804} {10.1063/5.0038804} (\bibinfo {year}
  {2021})\BibitemShut {NoStop}%
\bibitem [{\citenamefont {Wilczek}(1987)}]{Wilczek1987}%
  \BibitemOpen
  \bibfield  {author} {\bibinfo {author} {\bibfnamefont {F.}~\bibnamefont
  {Wilczek}},\ }\bibfield  {title} {\bibinfo {title} {Two applications of axion
  electrodynamics},\ }\href {https://doi.org/10.1103/PhysRevLett.58.1799}
  {\bibfield  {journal} {\bibinfo  {journal} {Phys. Rev. Lett.}\ }\textbf
  {\bibinfo {volume} {58}},\ \bibinfo {pages} {1799} (\bibinfo {year}
  {1987})}\BibitemShut {NoStop}%
\bibitem [{\citenamefont {Taherinejad}\ and\ \citenamefont
  {Vanderbilt}(2015)}]{Taherinejad2015}%
  \BibitemOpen
  \bibfield  {author} {\bibinfo {author} {\bibfnamefont {M.}~\bibnamefont
  {Taherinejad}}\ and\ \bibinfo {author} {\bibfnamefont {D.}~\bibnamefont
  {Vanderbilt}},\ }\bibfield  {title} {\bibinfo {title} {{Adiabatic Pumping of
  Chern-Simons Axion Coupling}},\ }\href
  {https://doi.org/10.1103/PhysRevLett.114.096401} {\bibfield  {journal}
  {\bibinfo  {journal} {Phys. Rev. Lett.}\ }\textbf {\bibinfo {volume} {114}},\
  \bibinfo {pages} {096401} (\bibinfo {year} {2015})}\BibitemShut {NoStop}%
\bibitem [{\citenamefont {Olsen}\ \emph {et~al.}(2017)\citenamefont {Olsen},
  \citenamefont {Taherinejad}, \citenamefont {Vanderbilt},\ and\ \citenamefont
  {Souza}}]{Olsen2017}%
  \BibitemOpen
  \bibfield  {author} {\bibinfo {author} {\bibfnamefont {T.}~\bibnamefont
  {Olsen}}, \bibinfo {author} {\bibfnamefont {M.}~\bibnamefont {Taherinejad}},
  \bibinfo {author} {\bibfnamefont {D.}~\bibnamefont {Vanderbilt}},\ and\
  \bibinfo {author} {\bibfnamefont {I.}~\bibnamefont {Souza}},\ }\bibfield
  {title} {\bibinfo {title} {{Surface theorem for the Chern-Simons axion
  coupling}},\ }\href {https://doi.org/10.1103/PhysRevB.95.075137} {\bibfield
  {journal} {\bibinfo  {journal} {Phys. Rev. B}\ }\textbf {\bibinfo {volume}
  {95}},\ \bibinfo {pages} {075137} (\bibinfo {year} {2017})}\BibitemShut
  {NoStop}%
\bibitem [{\citenamefont {Rahav}\ \emph {et~al.}(2003)\citenamefont {Rahav},
  \citenamefont {Gilary},\ and\ \citenamefont {Fishman}}]{Rahav2003}%
  \BibitemOpen
  \bibfield  {author} {\bibinfo {author} {\bibfnamefont {S.}~\bibnamefont
  {Rahav}}, \bibinfo {author} {\bibfnamefont {I.}~\bibnamefont {Gilary}},\ and\
  \bibinfo {author} {\bibfnamefont {S.}~\bibnamefont {Fishman}},\ }\bibfield
  {title} {\bibinfo {title} {{Effective Hamiltonians for periodically driven
  systems}},\ }\href {https://doi.org/10.1103/PhysRevA.68.013820} {\bibfield
  {journal} {\bibinfo  {journal} {Phys. Rev. A}\ }\textbf {\bibinfo {volume}
  {68}},\ \bibinfo {pages} {013820} (\bibinfo {year} {2003})}\BibitemShut
  {NoStop}%
\bibitem [{\citenamefont {Goldman}\ and\ \citenamefont
  {Dalibard}(2014)}]{Goldman2014}%
  \BibitemOpen
  \bibfield  {author} {\bibinfo {author} {\bibfnamefont {N.}~\bibnamefont
  {Goldman}}\ and\ \bibinfo {author} {\bibfnamefont {J.}~\bibnamefont
  {Dalibard}},\ }\bibfield  {title} {\bibinfo {title} {{Periodically Driven
  Quantum Systems: Effective Hamiltonians and Engineered Gauge Fields}},\
  }\href {https://doi.org/10.1103/PhysRevX.4.031027} {\bibfield  {journal}
  {\bibinfo  {journal} {Phys. Rev. X}\ }\textbf {\bibinfo {volume} {4}},\
  \bibinfo {pages} {031027} (\bibinfo {year} {2014})}\BibitemShut {NoStop}%
\bibitem [{\citenamefont {Eckardt}\ and\ \citenamefont
  {Anisimovas}(2015)}]{Eckardt2015}%
  \BibitemOpen
  \bibfield  {author} {\bibinfo {author} {\bibfnamefont {A.}~\bibnamefont
  {Eckardt}}\ and\ \bibinfo {author} {\bibfnamefont {E.}~\bibnamefont
  {Anisimovas}},\ }\bibfield  {title} {\bibinfo {title} {{High-frequency
  approximation for periodically driven quantum systems from a Floquet-space
  perspective}},\ }\href {https://doi.org/10.1088/1367-2630/17/9/093039}
  {\bibfield  {journal} {\bibinfo  {journal} {New Journal of Physics}\ }\textbf
  {\bibinfo {volume} {17}},\ \bibinfo {pages} {093039} (\bibinfo {year}
  {2015})}\BibitemShut {NoStop}%
\bibitem [{\citenamefont {Marin~Bukov}\ and\ \citenamefont
  {Polkovnikov}(2015)}]{Bukov2015}%
  \BibitemOpen
  \bibfield  {author} {\bibinfo {author} {\bibfnamefont {L.~D.}\ \bibnamefont
  {Marin~Bukov}}\ and\ \bibinfo {author} {\bibfnamefont {A.}~\bibnamefont
  {Polkovnikov}},\ }\bibfield  {title} {\bibinfo {title} {{Universal
  high-frequency behavior of periodically driven systems: from dynamical
  stabilization to Floquet engineering}},\ }\href
  {https://doi.org/10.1080/00018732.2015.1055918} {\bibfield  {journal}
  {\bibinfo  {journal} {Advances in Physics}\ }\textbf {\bibinfo {volume}
  {64}},\ \bibinfo {pages} {139} (\bibinfo {year} {2015})}\BibitemShut
  {NoStop}%
\bibitem [{\citenamefont {Mondragon-Shem}\ \emph {et~al.}(2019)\citenamefont
  {Mondragon-Shem}, \citenamefont {Martin}, \citenamefont {Alexandradinata},\
  and\ \citenamefont {Cheng}}]{Mondragon2019}%
  \BibitemOpen
  \bibfield  {author} {\bibinfo {author} {\bibfnamefont {I.}~\bibnamefont
  {Mondragon-Shem}}, \bibinfo {author} {\bibfnamefont {I.}~\bibnamefont
  {Martin}}, \bibinfo {author} {\bibfnamefont {A.}~\bibnamefont
  {Alexandradinata}},\ and\ \bibinfo {author} {\bibfnamefont {M.}~\bibnamefont
  {Cheng}},\ }\bibfield  {title} {\bibinfo {title} {Quantized frequency-domain
  polarization of driven phases of matter},\ }\Eprint
  {https://arxiv.org/abs/1811.10632} {arXiv:1811.10632}  (\bibinfo {year}
  {2019})\BibitemShut {NoStop}%
\bibitem [{\citenamefont {King-Smith}\ and\ \citenamefont
  {Vanderbilt}(1993)}]{KingSmith1993}%
  \BibitemOpen
  \bibfield  {author} {\bibinfo {author} {\bibfnamefont {R.~D.}\ \bibnamefont
  {King-Smith}}\ and\ \bibinfo {author} {\bibfnamefont {D.}~\bibnamefont
  {Vanderbilt}},\ }\bibfield  {title} {\bibinfo {title} {{Theory of
  polarization of crystalline solids}},\ }\href
  {https://doi.org/10.1103/PhysRevB.47.1651} {\bibfield  {journal} {\bibinfo
  {journal} {Phys. Rev. B}\ }\textbf {\bibinfo {volume} {47}},\ \bibinfo
  {pages} {1651} (\bibinfo {year} {1993})}\BibitemShut {NoStop}%
\bibitem [{\citenamefont {Vanderbilt}(2018)}]{Vanderbilt2018}%
  \BibitemOpen
  \bibfield  {author} {\bibinfo {author} {\bibfnamefont {D.}~\bibnamefont
  {Vanderbilt}},\ }\href {https://doi.org/10.1017/9781316662205} {\emph
  {\bibinfo {title} {{Berry Phases in Electronic Structure Theory: Electric
  Polarization, Orbital Magnetization and Topological Insulators}}}}\ (\bibinfo
   {publisher} {Cambridge University Press},\ \bibinfo {address} {Cambridge},\
  \bibinfo {year} {2018})\BibitemShut {NoStop}%
\bibitem [{\citenamefont {Kitagawa}\ \emph {et~al.}(2010)\citenamefont
  {Kitagawa}, \citenamefont {Berg}, \citenamefont {Rudner},\ and\ \citenamefont
  {Demler}}]{Kitagawa2010}%
  \BibitemOpen
  \bibfield  {author} {\bibinfo {author} {\bibfnamefont {T.}~\bibnamefont
  {Kitagawa}}, \bibinfo {author} {\bibfnamefont {E.}~\bibnamefont {Berg}},
  \bibinfo {author} {\bibfnamefont {M.}~\bibnamefont {Rudner}},\ and\ \bibinfo
  {author} {\bibfnamefont {E.}~\bibnamefont {Demler}},\ }\bibfield  {title}
  {\bibinfo {title} {{Topological characterization of periodically driven
  quantum systems}},\ }\href {https://doi.org/10.1103/PhysRevB.82.235114}
  {\bibfield  {journal} {\bibinfo  {journal} {Phys. Rev. B}\ }\textbf {\bibinfo
  {volume} {82}},\ \bibinfo {pages} {235114} (\bibinfo {year}
  {2010})}\BibitemShut {NoStop}%
\bibitem [{\citenamefont {Stegmaier}\ \emph {et~al.}(2024)\citenamefont
  {Stegmaier}, \citenamefont {Fritzsche}, \citenamefont {Sorbello},
  \citenamefont {Greiter}, \citenamefont {Brand}, \citenamefont {Barko},
  \citenamefont {Hofer}, \citenamefont {Schwingenschlögl}, \citenamefont
  {Moessner}, \citenamefont {Lee}, \citenamefont {Szameit}, \citenamefont
  {Alu}, \citenamefont {Kießling},\ and\ \citenamefont
  {Thomale}}]{Stegmaier2024}%
  \BibitemOpen
  \bibfield  {author} {\bibinfo {author} {\bibfnamefont {A.}~\bibnamefont
  {Stegmaier}}, \bibinfo {author} {\bibfnamefont {A.}~\bibnamefont
  {Fritzsche}}, \bibinfo {author} {\bibfnamefont {R.}~\bibnamefont {Sorbello}},
  \bibinfo {author} {\bibfnamefont {M.}~\bibnamefont {Greiter}}, \bibinfo
  {author} {\bibfnamefont {H.}~\bibnamefont {Brand}}, \bibinfo {author}
  {\bibfnamefont {C.}~\bibnamefont {Barko}}, \bibinfo {author} {\bibfnamefont
  {M.}~\bibnamefont {Hofer}}, \bibinfo {author} {\bibfnamefont
  {U.}~\bibnamefont {Schwingenschlögl}}, \bibinfo {author} {\bibfnamefont
  {R.}~\bibnamefont {Moessner}}, \bibinfo {author} {\bibfnamefont {C.~H.}\
  \bibnamefont {Lee}}, \bibinfo {author} {\bibfnamefont {A.}~\bibnamefont
  {Szameit}}, \bibinfo {author} {\bibfnamefont {A.}~\bibnamefont {Alu}},
  \bibinfo {author} {\bibfnamefont {T.}~\bibnamefont {Kießling}},\ and\
  \bibinfo {author} {\bibfnamefont {R.}~\bibnamefont {Thomale}},\ }\bibfield
  {title} {\bibinfo {title} {{Topological Edge State Nucleation in Frequency
  Space and its Realization with Floquet Electrical Circuits}},\ }\Eprint
  {https://arxiv.org/abs/2407.10191} {arXiv:2407.10191}  (\bibinfo {year}
  {2024})\BibitemShut {NoStop}%
\bibitem [{\citenamefont {Zhang}\ \emph {et~al.}(2025)\citenamefont {Zhang},
  \citenamefont {Cao}, \citenamefont {Qian}, \citenamefont {Yuan},\ and\
  \citenamefont {Zhang}}]{Zhang2025}%
  \BibitemOpen
  \bibfield  {author} {\bibinfo {author} {\bibfnamefont {W.}~\bibnamefont
  {Zhang}}, \bibinfo {author} {\bibfnamefont {W.}~\bibnamefont {Cao}}, \bibinfo
  {author} {\bibfnamefont {L.}~\bibnamefont {Qian}}, \bibinfo {author}
  {\bibfnamefont {H.}~\bibnamefont {Yuan}},\ and\ \bibinfo {author}
  {\bibfnamefont {X.}~\bibnamefont {Zhang}},\ }\bibfield  {title} {\bibinfo
  {title} {{Topolectrical space-time circuits}},\ }\bibfield  {journal}
  {\bibinfo  {journal} {Nature Communications}\ }\textbf {\bibinfo {volume}
  {16}},\ \href {https://doi.org/10.1038/s41467-024-55425-1}
  {10.1038/s41467-024-55425-1} (\bibinfo {year} {2025})\BibitemShut {NoStop}%
\bibitem [{\citenamefont {Kishony}\ \emph {et~al.}(2025)\citenamefont
  {Kishony}, \citenamefont {Grossman}, \citenamefont {Lindner}, \citenamefont
  {Rudner},\ and\ \citenamefont {Berg}}]{Kishony2025}%
  \BibitemOpen
  \bibfield  {author} {\bibinfo {author} {\bibfnamefont {G.}~\bibnamefont
  {Kishony}}, \bibinfo {author} {\bibfnamefont {O.}~\bibnamefont {Grossman}},
  \bibinfo {author} {\bibfnamefont {N.}~\bibnamefont {Lindner}}, \bibinfo
  {author} {\bibfnamefont {M.}~\bibnamefont {Rudner}},\ and\ \bibinfo {author}
  {\bibfnamefont {E.}~\bibnamefont {Berg}},\ }\bibfield  {title} {\bibinfo
  {title} {Topological excitations at time vortices in periodically driven
  systems},\ }\href {https://doi.org/10.1038/s41535-025-00745-8} {\bibfield
  {journal} {\bibinfo  {journal} {npj Quantum Materials}\ }\textbf {\bibinfo
  {volume} {10}},\ \bibinfo {pages} {28} (\bibinfo {year} {2025})}\BibitemShut
  {NoStop}%
\end{thebibliography}
\end{document}